\documentclass[manuscript]{aastex61}
\usepackage{graphicx}
\usepackage{amssymb}

\newcommand{\kms}{km\,s$^{-1}$}
\newcommand{\ho}{km\,s$^{-1}$\,Mpc$^{-1}$}
\newcommand{\ha}{\textrm{H}$\alpha$}
\newcommand{\Msun}{M$_{\odot}$}

\submitjournal{ApJ}
\shorttitle{TRGB distances to galaxies in front of Virgo}
\shortauthors{Karachentsev et al.}

\begin{document}

\title{TRGB distances to galaxies in front of the Virgo cluster}

\correspondingauthor{Igor Karachentsev}

\email{ikar@sao.ru}

\author{Igor D. Karachentsev}
\affiliation{Special Astrophysical Observatory, Nizhniy Arkhyz, 
Karachai-Cherkessia 369167, Russia}

\author{Lidia N. Makarova}
\affiliation{Special Astrophysical Observatory, Nizhniy Arkhyz, 
Karachai-Cherkessia 369167, Russia}

\author{R. Brent Tully}
\affiliation{Institute for Astronomy, University of Hawaii, 2680 Woodlawn Drive, 
HI 96822, USA}

\author{Luca Rizzi}
\affiliation{W. M. Keck Observatory, 65-1120 Mamalahoa Hwy, Kamuela, HI 96743, USA}

\author{Edward J. Shaya}
\affiliation{Astronomy Department, University of Maryland, College Park, MD 20743, USA}

\begin{abstract}

Tip of the red giant branch distances are acquired from Hubble Space Telescope images for 16 
galaxies to the foreground of the Virgo Cluster.  The new distances with 5\% accuracy, combined 
with archival measurements, tightly constrain the near side location of the onset of infall into 
the Virgo Cluster to be $7.3\pm0.3$~Mpc from the cluster, reaching within 9 Mpc of the Milky
Way. The mass within this turnaround radius about the cluster is $8.3\pm0.9 \times 10^{14}$~\Msun.
Color-magnitude diagrams are provided for galaxies in the study and there is brief discussion 
of their group affiliations. 

\end{abstract}


\section{Introduction}
The nearest galaxies lie in a flattened structure on the supergalactic equator 
that we have called the Local Sheet \citep{tully2008}. The Local Sheet is a 
wall of the Local Void that occupies most of the nearby volume above the 
supergalactic equator (positive SGZ). The Virgo Cluster lies within the 
supergalactic equatorial band beyond the Local Sheet at 16 Mpc.

 Over the years, imaging with Hubble Space Telescope (HST) has led to estimates 
of the distances to nearby galaxies from the measured luminosities of the 
brightest resolved red giant branch stars: the so-called tip of the red giant 
branch (TRGB) method. These observations have culminated in two HST-SNAP programs 
which gave, for any galaxy within 10 Mpc, a distance accurate to 5\% with a 
single HST orbit. There are now almost 400 accurate TRGB distances in the Local 
Volume Galaxy Database ( http://www.sao.ru/lv/lvgdb/) and Extragalactic 
Distance Database (EDD) (http://edd.ifa.hawaii.edu/). The morphology and kinematics 
of nearby structure is becoming increasingly clarified. On a very local scale 
($<$5 Mpc) we now have a detailed knowledge of the nature of galaxy groups: 
their memberships, dimensions, velocity dispersions, and infall patterns 
\citep{kar2005,tully2015,kash2018,shaya2017}. 
Associations of dwarfs probe the clustering regime at the halo level 
of $10^{11} M_{\sun}$ \citep{tullyetal2006}. The curious phenomenon of satellite 
layering is being documented \citep{ibata2013,shaya2013,tullyetal2015}. 

There are gaps between the Local Sheet and neighboring structure. The galaxies 
within the Local Sheet are found to be moving coherently, with minimal dispersion. 
The distances to galaxies within the Local Sheet are very close to expectations 
from the Hubble flow \citep{kar2009}. By contrast, nearby galaxies 
off the supergalactic equator essentially all have substantial negative 
velocities compared to Hubble law expectations. The structures below the plane 
that are accessible to TRGB measurements were given the names Leo Spur and 
Dorado Cloud in the Nearby Galaxies Atlas \citep{tullyfisher87}. There is a 
clear velocity discontinuity between the Local Sheet and these structures. The 
discontinuity is explained by a bulk motion of the Local Sheet toward negative 
SGZ of 260 \kms{} due to the expansion of the Local Void \citep{tully2008,kar2015}, 
a motion not experienced by the filaments at lower 
SGZ. In the opposite direction toward positive SGZ there are few galaxies and 
only two in the Local Void with TRGB distances: KK246 and ALFAZOAJ1952+1428
\citep{rizzi2017}. Both these objects have large peculiar velocities toward the Local Sheet 
and away from the void center.

In addition to the motion away from the Local Void, members of the Local Sheet 
are moving together toward the Virgo Cluster, roughly in the positive SGY 
direction, at  185 \kms{} \citep{aaronson82,tully2008}. 

This increasingly detailed knowledge of our local region speaks to the success 
of the HST-SNAP programs. However, the SNAP approach has reached a limitation. 
Most galaxies within 10 Mpc, the limit of the TRGB capability with single orbit 
HST observations, lie close to the supergalactic equator, indeed, in the Local 
Sheet, and pile up in a narrow Right Ascension window around 12 hours. The SNAP 
programs have led to coverage of a high fraction of available targets around
the sky except in the RA = 12 - 13 hours window. There has been a poor understanding 
of the nature of structure and kinematics of the Local Sheet in a swath of sky 
between the Canes Venatici II region and the Sombrero 
galaxy in the distance range 5-10 Mpc. While most galaxies within 4 Mpc have 
established TRGB distances, most galaxies in the Local Sheet that are suspected 
to lie in the interval 4-8 Mpc have not had accurately known distances. 

This region is particularly interesting, because it borders the Virgo Cluster 
infall domain. In the spherical approximation, it has been estimated that the cluster 
turnaround radius 
separating infall from expansion lies $7.2\pm0.7$ Mpc from the Virgo core \citep{kar2014}.
The front edge of this surface lies 9 Mpc from us. 
The volume that the Lambda-CDM model predicts will ultimately collapse into the
Virgo Cluster \citep{peirani2008} extends to within ~ 6 Mpc of 
our location. In this radial domain the density of galaxies associated with the Local Sheet falls 
off rapidly. Galaxies toward the far edge of the Local Sheet toward Virgo have increasingly high peculiar 
velocities away from us. Evidently, the attraction of the Virgo Cluster is
creating a velocity shear pattern.

Our proximity to the Virgo Cluster gives us a singular opportunity to study 
cluster infall, which carries information about the cluster mass distribution 
to much larger radii than most other measures (only weak lensing statistically 
explores comparable scales \citep{baum2006,uitert2011,bahcall2014,ziparo2016}). 
A plot from the \citet{kar2014} Virgo 
infall study shows that deviations from cosmic expansion evidently start to 
become substantial beyond 5 Mpc. Of course, the dynamics are in the non-linear 
regime but amenable to modeling in the spherical approximation. \citet{shaya2017}
show that the mass determined when using the spherical approximation is quite
accurate at this turnaround radius despite the components of non-radial flow.
A more rigorous analysis will be carried out that include all of these new high
accuracy galaxy measurements with numerical action methods 
\citep{peebles89,shaya95,peebles2001,shaya2013,shaya2017} in a future study.
The methodology is particularly well 
suited to the Virgo infall problem because, although the relationship between 
gravity and velocity is non-linear, the orbits have not reached shell-crossing 
so have only modest curvature.

In this work we present distances via the optical TRGB method for 16
galaxies in front of the Virgo Cluster using their imaging from HST. In
Section 2 we outline selection of targets, results of stellar photometry,
determination of galaxy distances, and summarize some individual properties
of the galaxies. In Section 3 we use the new accurate distances, combined
with radial velocities from NED (http://ned.ipac.caltech.edu)and additional 
distances from EDD, to constrain the
near-side of the Virgo cluster infall. Two approaches are applied to estimate
the uncertainty due to not knowing the true 3D motion of galaxies. As a result,
we tightly constrain the near side location of the infall zone around the
Virgo Cluster and estimate its total mass within the turnaround radius.
Finally, in Section 4 we draw our conclusions and give a brief outlook.

\section{HST/ACS observations}
\subsection{Selection of targets}

\begin{figure*}
\includegraphics[width=15cm]{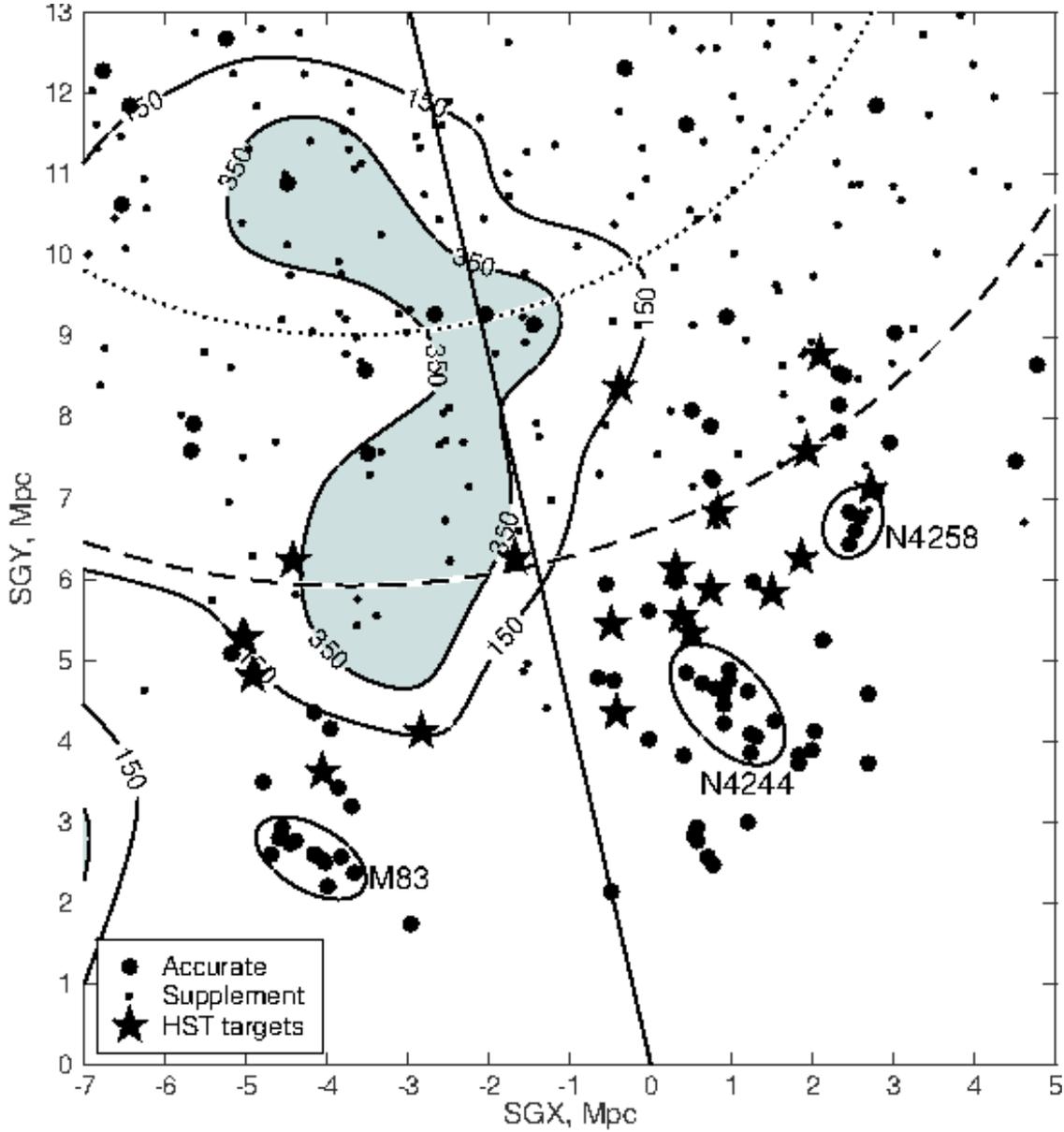}
\caption{ Distribution of galaxies in front of Virgo Cluster in Cartesian 
supergalactic coordinates. Solid circles and black dots are galaxies with 
accurate and supplement distances, respectively. The target objects are
shown by asterisks. The dotted and dashed large circles correspond to the
zero-velocity and zero-energy surface respectively. The region of high
positive peculiar velocities is shown in gray.}

\label{fig:SGBL}
\end{figure*}

Figure~\ref{fig:SGBL} presents a $SGB=\pm10\degr$ thick and $SGL=\pm40\degr$ wide wedge  
of the Local Sheet centered on M87 (SGB=$-2.35\degr$, SGL=$102.88\degr$) as the 
Virgo Cluster center. The Milky Way is at the origin. Galaxies with accurate 
TRGB or Cepheid distances are shown by solid circles and other cases with 
typically 20-25 per cent errors are indicated by black dots. The dotted ring 
identifies the projected zero velocity surface separating infall from cosmic 
expansion around the Virgo Cluster (SGX=$-3.7$ Mpc, SGY=$16.1$ Mpc) in the 
spherical approximation. The straight line shows the direction toward M87. 
The dashed circle corresponds roughly to the zero energy surface, the
outer extreme of the region that will ultimately fall into the Virgo Cluster 
assuming the current LCDM paradigm. Positions of three nearby groups: M~83, NGC~4244 
and NGC~4258 are marked. The gray area and the solid curve indicate a zone 
of positive peculiar velocities $V_{pec} = V_{LG}-73D_{Mpc} > 350$ \kms{} 
and $V_{pec}= 150 $ \kms{} after smoothing by a Gaussian 
filter with a window of 0.75 Mpc. There is an evident flow toward the Virgo 
Cluster. The asterisks identify 20 targets for our program of TRGB distance 
measurements at distances assigned from redshifts, Tully-Fisher relation and 
other secondary indicators. The objects are situated within $SGB=\pm10\degr$ 
and $SGL=\pm40\degr$ with respect to M87. Their expected distances lie in 
the range of 5-9 Mpc.
 
A list of targets is given in Table~\ref{table:list}, where the columns contain: (1,2) galaxy
name and its number in HyperLEDA \citep{makarov2014}, (3,4) equatorial (J2000)
and supergalactic coordinates, (5) total B-magnitude, (6) morphological type in
numerical code according to \citet{vauc1991}, (7) radial velocity \kms{}
in the Local Group reference frame adopted in NASA Extragalactic Database (NED),
(8,9) distance (Mpc) and the method of its estimate: from the linear Hubble flow 
with parameter $H_0 = 73$ \ho{} (h), from \citet{tully77} relation (TF), 
via membership in the known groups (mem); distances to two last objects: KDG218
and NGC5068 were estimated from dwarf galaxy texture (txt) and the Planetary Nebulae
luminosity function (PL). 

\begin{table*}
\centering
\caption{List of targets for GO 14636}
\label{table:list}
\begin{tabular}{lrcccrcclcc}
\hline
   Name   &   LEDA &    RA   Dec &   SGL   SGB & $B_T$ &  T & $V_{LG}$ & D & meth  \\
\hline
   (1)    &   (2)  &      (3)    &     (4)     &  (5)  &(6) & (7)    &(8) & (9)  \\
\hline
BTS76         & 2832100& 115844.1+273506 & 86.06$-$04.71 &16.5 & 10 & 407 & 5.57 &h \\
LV J1217+3231 & 4319966& 121732.0+323157 & 82.75+00.77 &18.2 &  9 & 433 & 5.93 &h \\
LV J1218+4655 & 4320422& 121811.1+465501 & 69.07+05.31 &16.8 &  8 & 477 & 7.66 &mem \\
KK144         &  166137& 122527.9+282857 & 87.13+01.17 &16.5 & 10 & 449 & 6.15 &TF  \\
NGC4455       &  041066& 122844.1+224921 & 92.75+00.20 &12.9 &  7 & 588 & 8.40 &TF \\
KK151         &  041314& 123023.8+425405 & 73.56+06.24 &15.8 &  9 & 479 & 6.56 &TF \\
UGC07678      &  041522& 123200.4+394955 & 76.61+05.70 &13.8 &  9 & 710 & 9.08 &TF \\
KK152         &  041701& 123324.9+332105 & 82.95+04.20 &16.3 & 10 & 836 & 6.90 &TF  \\
KKSG29        &  042120& 123714.1$-$102951 &125.42$-$07.05 &16.5 & 10 & 562 & 7.70 &TF \\
NGC4618       &  042575& 124132.8+410903 & 75.80+07.79 &11.3 &  6 & 576 & 7.90 &TF \\
LV J1243+4127 & 5056993& 124355.7+412725 & 75.61+08.30 &17.2 & 10 & 444 & 6.09 &h  \\
NGC4656       &  042863& 124357.6+321013 & 84.68+06.02 &11.0 &  8 & 635 & 5.40 &TF \\
IC3840        &  086644& 125146.1+214407 & 95.26+05.03 &16.9 & 10 & 510 & 5.50 &TF \\
UGC08061      &  044170& 125643.4+115552 &105.06+03.61 &16.0 & 10 & 474 & 6.49 &h  \\
KK176         &  044681& 125956.3$-$192447 &135.49$-$04.04 &16.5 & 10 & 618 & 6.90 &TF \\
UGCA319       &  044982& 130214.4$-$171415 &133.54$-$02.94 &15.1 & 10 & 555 & 7.30 &mem \\
KK177         &  087149& 130241.9+215951 & 95.64+07.55 &17.4 & $-$2 & 228 & 4.40 &mem \\
DDO161        &  045084& 130316.8$-$172523 &133.78$-$02.75 &13.5 &  8 & 543 & 7.30 &TF \\
KDG218        &  045303& 130544.0$-$074520 &124.64+00.46 &16.8 & $-1$ &  $-$  & 5.00 &txt \\
NGC5068       &  046400& 131855.3$-$210221 &138.28$-$00.21 &10.5 &  6 & 469 & 5.45 &PL \\
\hline
\end{tabular}
\end{table*}

\subsection{Observations and photometry}
Observations of the 20 galaxies were made with the ACS/HST
within the project GO-14636 (PI: I.D.Karachentsev). 
For each galaxy, one exposure was obtained in the F606W filter 
and one exposure in the F814W filter with the exposure time 1030~s in each filter.
Figure~\ref{fig:ima} shows
the F606W images of the galaxies. Most galaxies are
well resolved into individual stars including the
red giant branch (RGB).

\begin{figure*}
\centerline{\includegraphics[width=7cm]{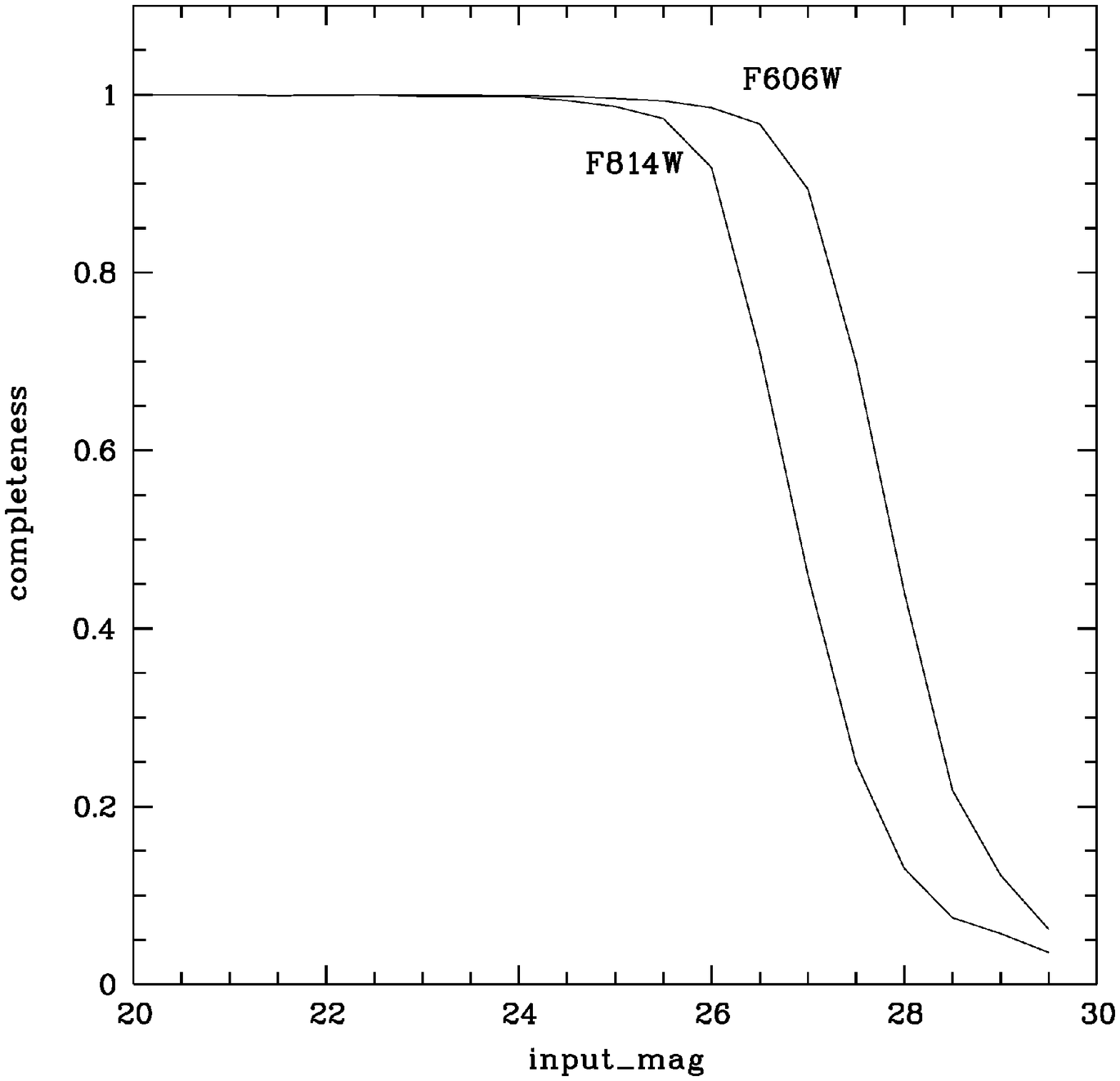}}
\caption{Photometric completeness as a function of magnitude for KK152, which is an example 
of fainter TRGB magnitude in our sample. As can be seen from the figure, even for this case
the level of the completeness is high, and does not constrain our distance measurement. }
\label{fig:compl}
\end{figure*}

\begin{figure*}
\includegraphics[width=15cm]{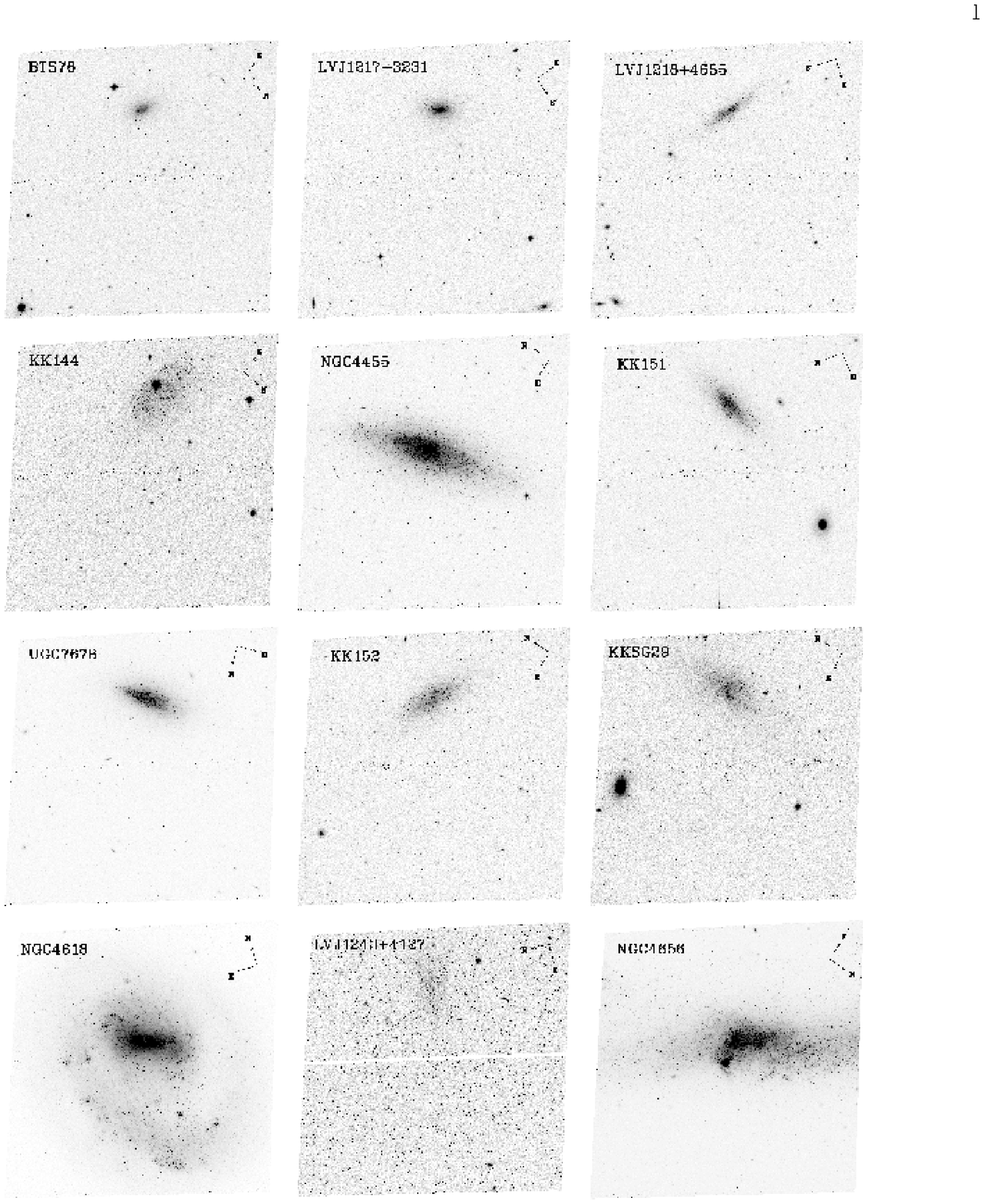}
\caption{\textit{HST}/ACS combined distortion-corrected  mosaic 
image of the target galaxies in the \textit{F606W} filter. 
The image size is $3.4\times3.4$ arcmin.
}
\label{fig:ima}
\end{figure*}

\addtocounter{figure}{-1}
\begin{figure*}
\includegraphics[width=15cm]{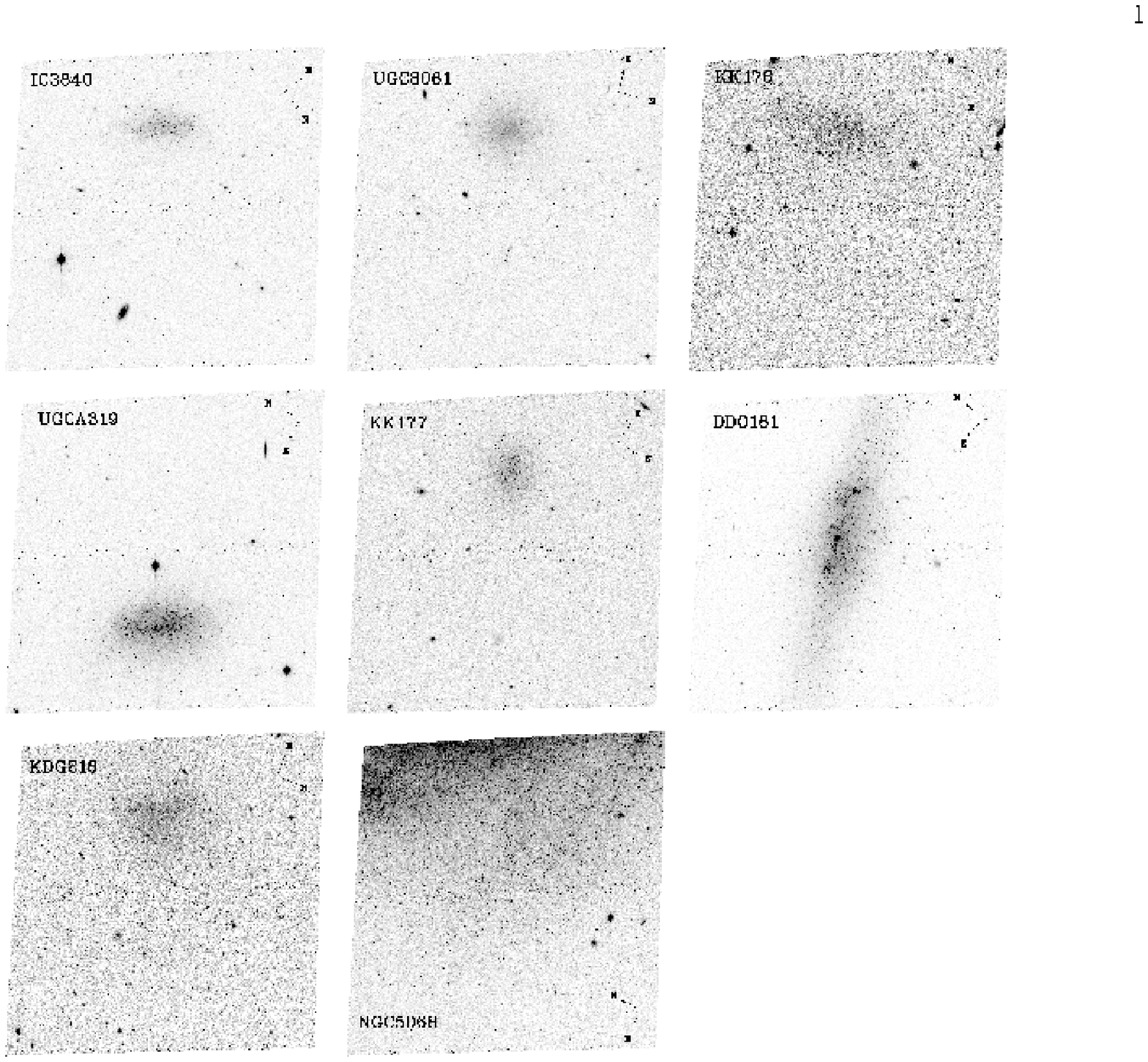}
\caption{Continued.}
\label{fig:ima}
\end{figure*}

We use the ACS module of the {\sc DOLPHOT} software package\footnote{http://americano.dolphinsim.com/dolphot/}
by A. Dolphin \citep{d2016} for photometry of resolved stars.
The data quality images were used to mask bad pixels. 
Only stars of good photometric quality were used in the
analysis, following the recommendations given in the {\sc DOLPHOT} User's 
Guide. We have selected the stars with signal-to-noise (S/N) of at least 
five in both filters and $\vert sharp \vert \le 0.3$.
The resulting color-magnitude diagrams of the galaxies are presented in Fig.~\ref{fig:cmd}.

Artificial star tests provide an accurate way 
to estimate the photometric errors, blending and 
incompleteness in the crowded fields of nearby resolved galaxies.
These tests were performed for the objects under study using the same  {\sc DOLPHOT}
reduction procedures. 
Sufficient artificial stars were generated 
in the appropriate magnitude and color range so that
the distribution of the recovered magnitudes is adequately sampled. 
Photometric errors are indicated in Fig.~\ref{fig:cmd}. In the CMDs we show stars 
with signal-to-noise $\ge$5 in both F606W and F814W filters.
An example of the completeness function is given in Fig.~\ref{fig:compl} for KK152, which
has one of the faintest TRGB value. As can be seen from the figure, even for this case
the level of the completeness is high, and does not constrain our distance measurement.
For most objects only stars located within the body of the galaxy are shown and used in further analysis. 
In the few cases when the galaxy is large and bright, with a significant concentration 
of stars in the body, we use (and show in the CMD) only the stars located in the outer, minimally 
obscured  parts of this galaxy. These cases are indicated in the Section 2.4.

\subsection{Distance determination}

\begin{figure*}
\includegraphics[width=15cm]{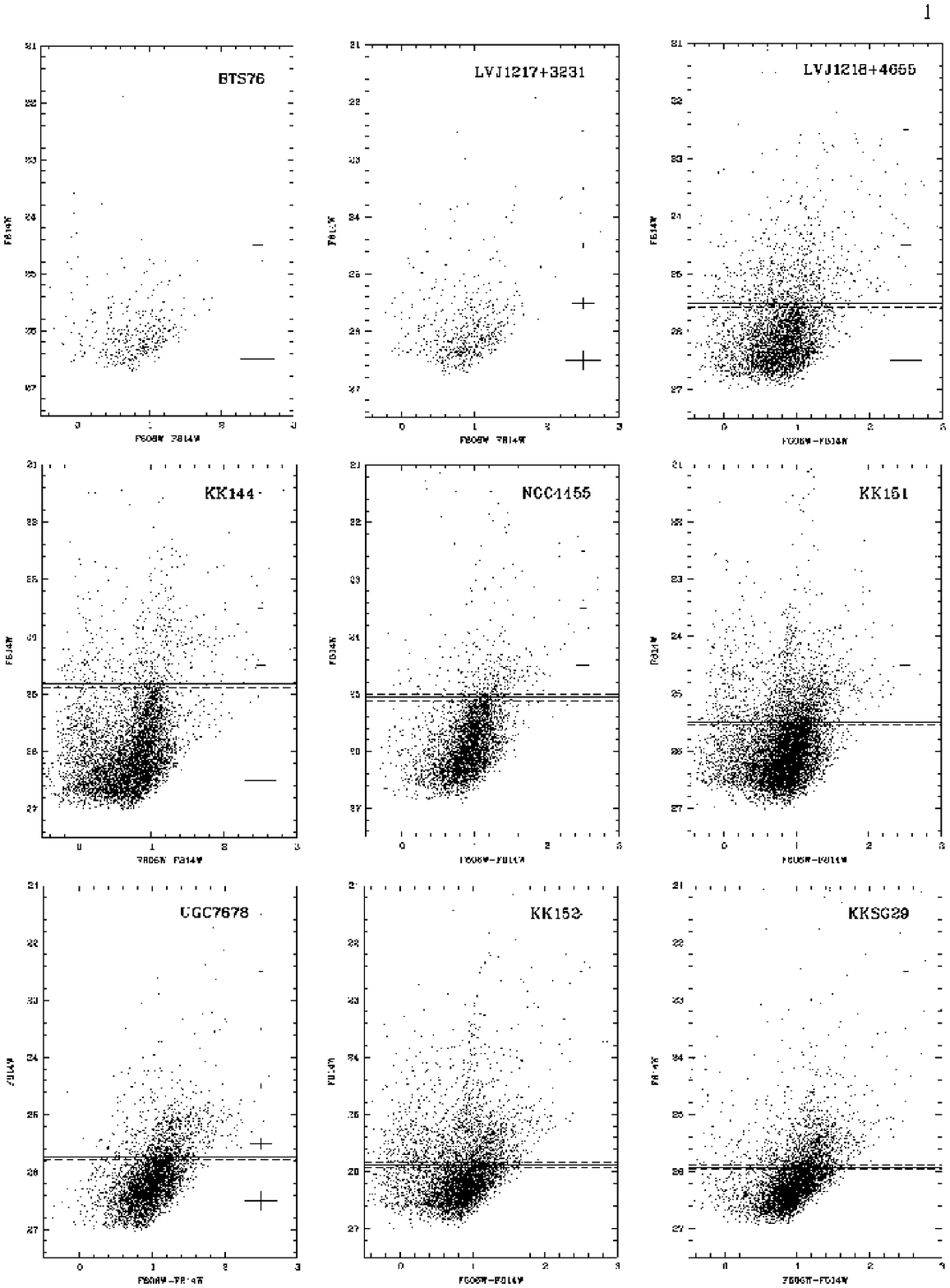}
\caption{{\footnotesize Color-magnitude diagrams of the target galaxies. 
Photometric errors are indicated by bars at the right in the CMD.
Derived TRGB values are indicated with solid line, whereas the TRGB
uncertainties are dashed lines.}
}
\label{fig:cmd}
\end{figure*}

\addtocounter{figure}{-1}
\begin{figure*}
\includegraphics[width=15cm]{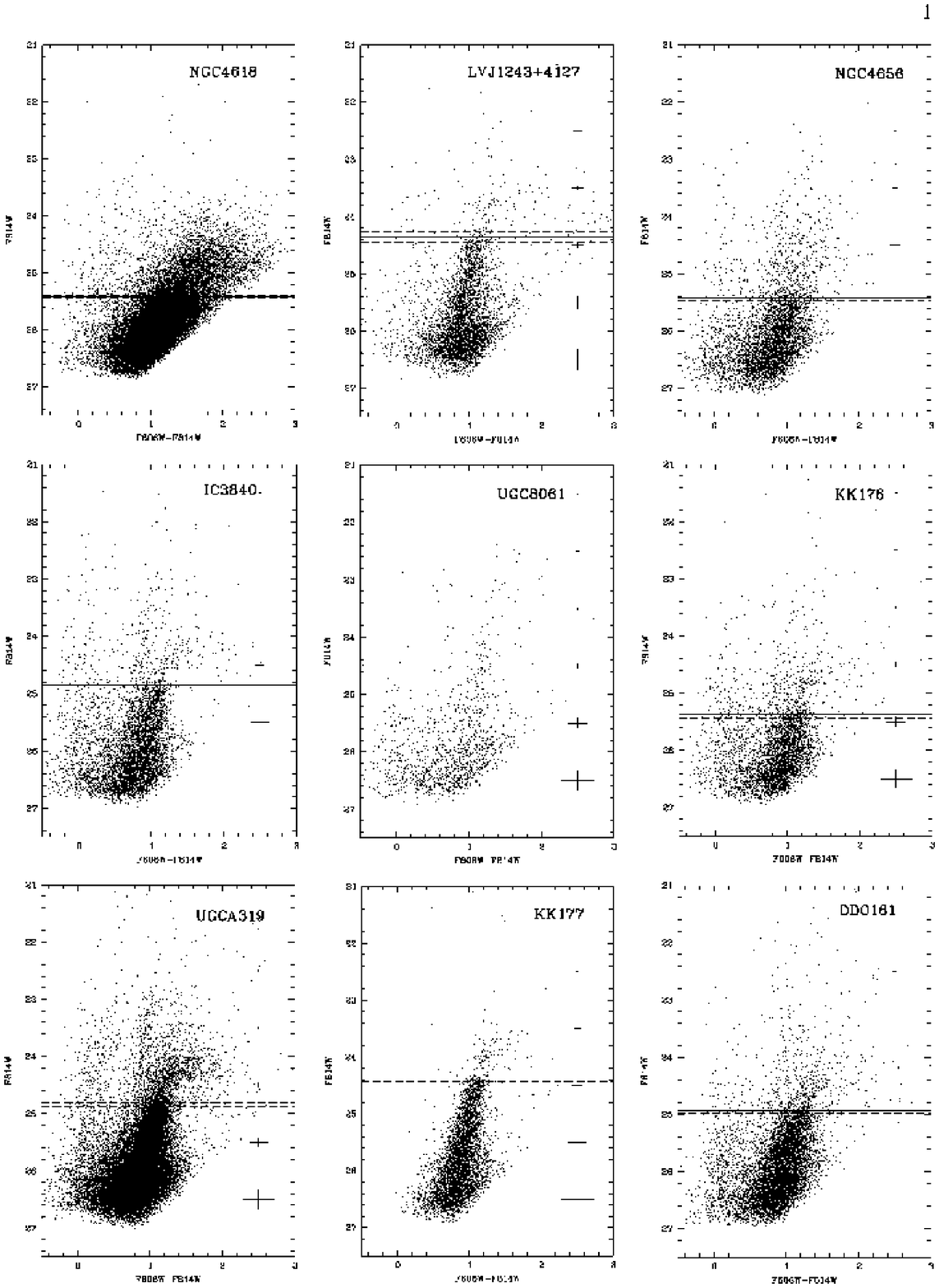}
\caption{Continued.
}
\label{fig:cmd}
\end{figure*}

\addtocounter{figure}{-1}
\begin{figure*}
\includegraphics[width=15cm]{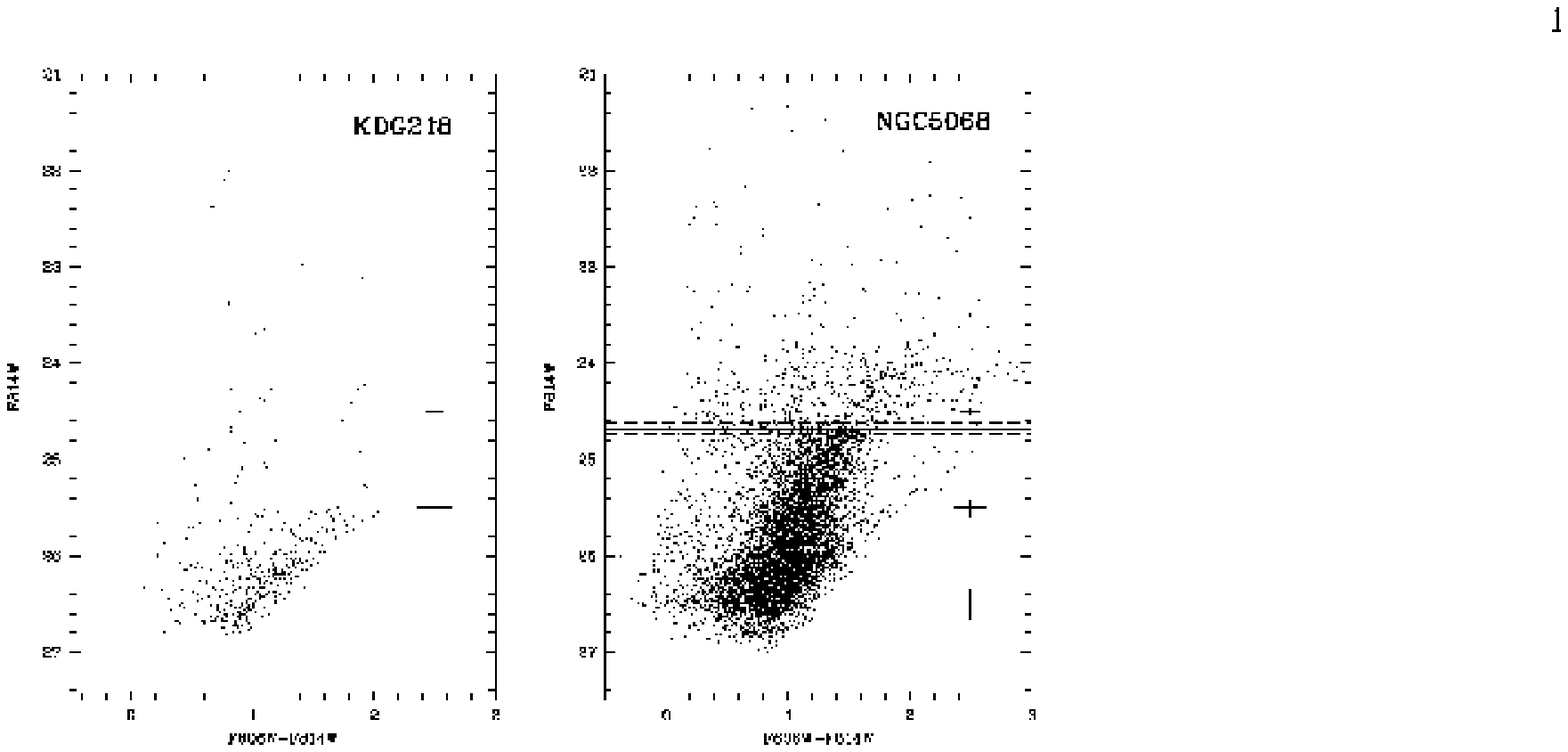}
\caption{Continued.
}
\label{fig:cmd}
\end{figure*}

We have determined the photometric TRGB distance with
our {\it trgbtool} program \citep{makarov06}.
The program uses a maximum-likelihood algorithm to fit the RGB stellar luminosity function
with a simple power law with a cutoff in the TRGB region plus a power law of
a second slope for a stellar population brighter than the TRGB.
The completeness, the photometric uncertainty, and the photometric bias  
are determined as a function of magnitude using artificial star experiments. 
They are incorporated in the maximum-likelihood fit to obtain a magnitude of TRGB. 
We have determined a respective $(F606W-F814W)_{TRGB}$ color 
using running mean over the RGB in the color-magnitude diagram of the galaxy.
The measured TRGB magnitudes in the ACS instrumental system are shown in the Table~\ref{table:dist}. 
Using the calibration for the TRGB distance indicator by 
\citet{rizzietal07} and the Galactic extinction from \citet{schlafly}, 
we derived the true distance moduli for the program galaxies (see Table~\ref{table:dist}).
The calibration of \citet{rizzietal07} includes a zero-point and color coefficients of the transformation.
The table includes: (1) galaxy name, (2) measured $F814W_{TRGB}$ value with the 68\% confidence 
level uncertainty
estimation, where photometric uncertainties are also included in the course of the maximum-likelihood
fitting, (3) measured $(F606W-F814W)_{TRGB}$ value with the uncertainty, estimated from
the running mean, (4) the Galactic extinction, (5) distance modulus with the uncertainty estimation,
including $F814W_{TRGB}$ and $(F606W-F814W)_{TRGB}$ value uncertainties, the Galactic extinction uncertainty and the uncertainties of \citet{rizzietal07} calibration, including the color term calibration uncertainty.

\begin{table*}
\centering
\caption{TRGB and distance measurements for the GO 14636 galaxies}
\label{table:dist}
\begin{tabular}{lccccc}
\hline
   Name   & $F814W_{TRGB}$ & $(F606W-F814W)_{TRGB}$ & $E(B-V)$ & $\mu_0$ & $D$, Mpc   \\
\hline
BTS76         &$>26.5$ & &$0.020$ &$>30.5$ & $>13$ \\
LV J1217+3231 &$>26.5$ & &$0.010$ &$>30.5$ & $>13$ \\
LV J1218+4655 & $25.50\pm0.07$ & $0.95_{-0.05}^{+0.02}$ & $0.017$ & $29.59\pm0.09$ & $8.26\pm0.35$ \\
KK144         & $24.82\pm0.06$ & $0.99_{-0.02}^{+0.03}$ & $0.021$ & $28.89\pm0.08$ & $5.99_{-0.22}^{+0.23}$ \\
NGC4455       & $25.06\pm0.06$ & $1.13_{-0.04}^{+0.02}$ & $0.019$ & $29.10\pm0.09$ & $6.61\pm0.27$ \\
KK151         & $25.50\pm0.04$ & $1.03\pm0.01$ & $0.018$ & $29.57\pm0.07$ & $8.20\pm0.26$ \\
UGC07678      & $25.74\pm0.04$ & $1.16\pm0.02$ & $0.014$ & $29.79\pm0.07$ & $9.08\pm0.30$ \\
KK152         & $25.88\pm0.05$ & $1.06_{-0.01}^{+0.02}$ & $0.016$ & $29.95\pm0.08$ & $9.77\pm0.35$ \\
KKSG29        & $25.92\pm0.04$ & $1.13\pm0.02$ & $0.026$ & $29.96\pm0.07$ & $9.81\pm0.31$ \\
NGC4618       & $25.42\pm0.02$ & $1.34\pm0.01$ & $0.019$ & $29.42\pm0.06$ & $7.66\pm0.22$ \\
LV J1243+4127 & $24.36\pm0.09$ & $1.15_{-0.02}^{+0.03}$ & $0.015$ & $28.41\pm0.11$ & $4.82\pm0.24$ \\
NGC4656       & $25.43\pm0.04$ & $1.04\pm0.01$ & $0.011$ & $29.51\pm0.07$ & $7.96\pm0.25$  \\
IC3840        & $24.85\pm0.06$ & $1.05\pm0.02$ & $0.035$ & $28.88\pm0.08$ & $5.99\pm0.23$  \\
UGC08061      &$>26.5$ & &$0.025$ &$>30.5$ &$>13$ \\
KK176         & $25.37\pm0.06$ & $1.08\pm0.01$ & $0.086$ & $29.31\pm0.09$ & $7.28\pm0.29$  \\
UGCA319       & $24.84\pm0.03$ & $1.01\pm0.01$ & $0.072$ & $28.80\pm0.07$ & $5.75\pm0.18$  \\
KK177         & $24.39\pm0.04$ & $1.10\pm0.01$ & $0.033$ & $28.42\pm0.07$ & $4.82\pm0.16$  \\
DDO161        & $24.94_{-0.04}^{+0.08}$ & $1.07\pm0.01$ & $0.070$ & $28.90_{-0.07}^{+0.10}$ & $6.03_{-0.21}^{+0.29}$  \\
KDG218        &$>26.5$ & &$0.040$ &$>30.5$ &$>13$ \\
NGC5068       & $24.68_{-0.06}^{+0.05}$ & $1.31_{-0.02}^{+0.04}$ & $0.091$ & $28.56_{-0.09}^{+0.08}$ & $5.16_{-0.21}^{+0.20}$  \\
\hline
\end{tabular}
\end{table*}

\subsection{Some individual properties of the galaxies}
{\bf BTS76 and LV J1217+3231.}  These two dwarf irregular (dIr) galaxies
have radial velocities provided by the Sloan Digital Sky Survey (SDSS) \citep{aba2009}.
Their distances were unknown. On the images
obtained with HST/ACS both galaxies are barely resolved into stars without
signs of the RGB. We infer that $I_{TRGB} > 26.5$ mag and distances $D > 13$ Mpc.
Judging by their low radial velocities, $V_{LG} < 500$ \kms{}, and distances $D > 13$ Mpc,
the galaxies are related to a mysterious association of 12 to 20 galaxies projected near the
E-galaxy NGC 4278 (Coma I Group) with peculiar velocities as extreme as 
$-1000$ \kms{} \citep{kar2011}. 

{\bf LV J1218+4655.} This dwarf spiral (Sm) galaxy resides at a projected separation 
of 25' or 56 kpc from the massive spiral NGC 4258, with a distance of 
$7.83\pm0.17$ Mpc determined by Cepheids \citep{newman2001}. Our
distance to LV\,J1218+4655 of $D = 8.26\pm0.35$~Mpc is compatible with the  
status of LV\,J1218+4655 as a companion to NGC 4258.

{\bf KK144.} This is an isolated low surface brightness (LSB) dIr galaxy with 
the high HI-mass-to-stellar-mass ratio $M_{HI}/M_*\sim4$.

{\bf NGC 4455 = EVCC0711.} This is a dwarf Sdm galaxy foreground to the Virgo Cluster.
Together with UGC~7584 and KKH~80 it probably forms an association of dwarfs having a 
projected radius of 30 kpc and velocity dispersion of 20 \kms{}.
We use in the distance measurement and show in the CMD only stars in the outer part of 
the galaxy. We selected the stars situated outside of the ellipse with the parameters: X=2133, 
Y=2439, A=1705, B=675, PA=73.8, where X,Y are the ellipse center pixel coordinates in the combined
distortion-corrected  mosaic F606W frame, A and B are major and minor axes in pix, and PA is 
position angle of the major axis in degrees.
 
{\bf MCG+07-26-012 = KK151.} This Magellanic-type (Im) dwarf is likely a 
peripheral companion to NGC~4258, with a projected separation of 660 kpc and radial velocity 
difference of 27 \kms{}.

{\bf UGC 7678.} This is a Im-type dwarf situated in vicinity of interacting pair
NGC~4485/4490 ( $D_{TRGB}= 8.91$ Mpc, EDD) at a projected separation of 300 kpc.
We use in the distance measurement and show in the CMD only stars in the outer part of 
the galaxy. We selected the stars situated outside of the ellipse with the parameters: X=2286, 
Y=3138, A=900, B=579, PA=15.

{\bf MCG+06-28-009 = KK152.} This is a gas-rich dIr galaxy. Together with
another dwarf galaxy AGC229089 and BCD galaxy NGC~4509 it probably forms a 
dwarf association with a projected radius of 90 kpc and velocity 
dispersion of 50 \kms{}. 

{\bf KKSG29.} This LSB dIr galaxy is a neighbor of NGC4594 ("Sombrero") at a 
projected separation of $1.3\degr$ or 220 kpc. However, their radial velocities
differ by 332 \kms{}.

{\bf NGC 4618.} This is a peculiar Sdm galaxy Arp 023 = VV073 with a distorted single-arm  
structure. Together with the nearby distorted Sdm galaxy NGC 4625 it
forms a pair KPG349 with a projected separation of 9' and radial velocity 
difference of 70 \kms{}. Surprisingly, the TRGB distance to NGC 4618, 
$7.66\pm0.22$ Mpc, differs significantly from that of NGC 4625,
$11.75\pm0.59$ Mpc \citep{mcquinn2017}. We re-measured the TRGB-distance to NGC 4625
using its archival HST images and obtained a distance of $12.61\pm0.49$ Mpc in
satisfactory agreement with the estimate by \citet{mcquinn2017}. This is an astonishing,
instructive case where a small angular separation between two galaxies, their 
small radial velocity difference, as well as apparent signs of interaction, 
do not ensure the spatial proximity of the galaxies.
We use in the distance measurement and show in the CMD only stars in the outer part of 
the galaxy. We selected the stars situated outside of the ellipse with the parameters: X=2242, 
Y=1994, A=2000, B=1560, PA=40.7.

{\bf LV J1243+4127.} This is a gas-rich LSB dIr galaxy with $M_{HI} \simeq M_*$. Its
projected separation from NGC 4736 (104 kpc) and small radial velocity
difference, 92 \kms{}, allows us to assign it to the NGC 4736 suite.

{\bf NGC 4656.} Hockey-stick-like Sm galaxy at a distance of $7.96\pm0.25$ Mpc is 
strongly distorted by interaction with the whale-like Sd galaxy NGC 4631
at $D = 7.38\pm0.23$ Mpc \citep{radburn2011}. Both the galaxies are
linked with a faint bridge \citep{kar2015a}. Near NGC 4656 on its
northern edge there is an LSB companion NGC 4656UV detected in ultraviolet
by GALEX.
The galaxy is large and does not fit in the ACS field. We selected in our measurements
the stars with Y$<$1560 in the pixel coordinates of the combined distortion-corrected  mosaic 
F606W frame.

{\bf IC 3840 = LSBC D575-01.} A gas-rich dIr galaxy with an emission \ha{}
semi-ring structure on the northern side.

{\bf UGC 8061 = EVCC1253.} This dIr galaxy lies $6.3\degr$ away from M87. Its
CMD does not show the presence of RGB above I = 26.5 mag, yielding a 
distance estimate $D > 13$ Mpc. Apparently, UGC8061 belongs to the virial
core of the Virgo Cluster.

{\bf KK176.} This is an isolated LSB dIr galaxy with $M_{HI}/M_*$ = 2.3. 
Its CMD has been presented by us earlier \citep{kar2017a}.

{\bf UGCA 319 and DDO161.} These are two gas-rich dwarf galaxies constituting a physical pair.
Their properties have been described recently by \citet{kar2017a}.
In the CMD and the distance measurement of DDO161 we selected
the stars with X$<$1476 or X$>$2772 in the pixel coordinates of the combined distortion-corrected  mosaic 
F606W frame.

{\bf KK177 = LSBC D575-03 = EVCC1290.} This is a LSB dwarf spheroidal galaxy. Judging
by its radial velocity of $V_{LG} = 228\pm60$ \kms{} \citep{kim2014} and distance 
of $D = 4.82\pm0.16$ Mpc, KK177 is likely a satellite of NGC 4826 having 
$V_{LG} = 365$ \kms{} and  $D = 4.41\pm0.20$ (EDD).

{\bf KDG218.} This is an ultra-diffuse dwarf galaxy associated with a group around NGC 4958
at $D = 22$ Mpc or with the Virgo Southern Extension filament at $D = 16$ Mpc.
Its properties and CMD has been discussed by \citet{kar2017b}.

{\bf NGC 5068.} The HST/ACS images are pointed on the SE periphery of this Scd galaxy
to avoid crowded HII-regions. The CMD of NGC 5068 has been presented earlier
\citep{kar2017a}. Here we selected in our measurements
the stars with Y$<$2066 in the pixel coordinates of the combined distortion-corrected  mosaic 
F606W frame. Our estimate of TRGB distance to it, $D = 5.16\pm0.21$ Mpc, is
well consistent with the PNLF distance $D = 5.45\pm0.52$ Mpc, derived by \citet{herr2008}. 
The galaxy is located on the outskirts of the group around NGC 5236 in Centaurus.

\section{Pattern of infall in front of Virgo Cluster} 
New accurate distance measurements are used by us to clarify the radius of zero-velocity 
surface of the Virgo Cluster. To this end, we calculated Virgo-centric distances
$$D_{VC}^2=D_g^2+D_c^2-2D_g\times D_c\times \cos \Theta,  $$ 
where $D_g$ and $D_c$ are distances to a galaxy and to the cluster center from observer, 
and $\Theta$ is the galaxy angular separation from M~87. The velocity of the galaxy with respect 
to the cluster center is determined by us in two models of "minor" and "major" attractors.
The first case assumes that peculiar velocities of galaxies in the cluster
vicinity are small in comparison with the regular Hubble flow, the second 
case implies that the infall velocity towards the cluster center prevails
for most the galaxies over their tangential motions (see details in
\citet{kn2010}).

In the model of "minor" attractor the Virgo-centric velocity is
 $$ V_{VC}=V_g\times \cos\lambda -V_c\cos(\lambda+\Theta),$$
where
$$\tan\lambda=D_c\times \sin\Theta/(D_g-D_c\times \cos\Theta), $$
and in the case of "major" attractor it is
$$V_{VC}=[V_c\times \cos\Theta -V_g]/\cos\lambda.$$
Here, $\lambda$ means the angle between the line of sight towards a galaxy 
and the line connecting the galaxy with the cluster center. 
This dichotomy is due to our ignorance of the 
full spatial velocity of the galaxy. The difference between the two models becomes 
negligible when a galaxy locates just in front of the cluster center 
$(\lambda\simeq180^{\degr})$. 

The terms $V_{VC}$ and $D_{VC}$ require the adoption of the distance
and the mean velocity of the Virgo center. Following \citet{binggeli93, mei2007},
and Kashibadze et al. (2018, in preparation), we adopt: $V_c = 984$ \kms{}, $D_c = 16.65$ Mpc
that yields the peculiar velocity of the Virgo cluster (i.e. the infall velocity of the LG towards Virgo) 
as $V_{c,pec} = -231$ \kms{}.

Among the galaxies with accurate distances in front of Virgo we selected 28
objects with a suitable angle $\lambda=[135-180^{\degr}]$. Pertinent data are
collected in the Table~\ref{table:virgogal}. The columns contain: (1) galaxy name, (2) supergalactic
coordinates, (3) radial velocity in the LG rest frame, (4) galaxy distance
from the MW, (5) angular separation from M~87, (6) Virgo-centric distance,
(7) angular separation between the line of sight and the direction from the galaxy
towards M~87, (8,9) Virgo-centric velocity in cases of the minor and the major
attractor, respectively, (10) reference for the distance source. The galaxies 
are ranked by their distance $D_{VC}$ in
a range of 3 - 13 Mpc, hence avoiding the Virgo core. As can be seen,
only eight of twenty of our targets satisfy the geometric condition
$\lambda = 135-180^{\circ}$. 

\begin{table*}
\centering
\caption{Galaxies with TRGB-distances in front of the Virgo Cluster.}
\label{table:virgogal}
\begin{tabular}{lrrrrrrrrl}
\hline
  Name   &   SGL   SGB &  $V_{LG}$ &  $D$  & $\Theta$ & $D_{VC}$ & $\lambda$ & $V_{VC}^{mi}$ & $V_{VC}^{ma}$ & Ref \\
         &      deg    &  \kms{}   & Mpc   &  deg  &     Mpc  & deg    &  \kms{}       & \kms{}     \\
\hline
  (1)    &      (2)    &  (3)      & (4)   &  (5)  &     (6)  &   (7)  &    (8)        & (9)  & (10)      \\
\hline
 UGC07512 & 112.36{-}6.46& 1353 & 11.86 &  10 &  5.41 & 147 &{-}225 &{-}461 & EDD \\
 VCC2037  & 106.03+0.67  & 1032 &  9.64 &   4 &  7.08 & 170 & {-}37 & {-}52 & EDD \\
 IC3583   & 102.45{-}0.72& 1024 &  9.51 &   2 &  7.15 & 176 & {-}38 & {-}41 & EDD \\
 GR34     &  99.00{-}3.36& 1204 &  9.29 &   4 &  7.41 & 171 &{-}209 &{-}225 & EDD \\
 NGC4600  & 112.45{-}2.67&  696 &  9.29 &  10 &  7.65 & 159 &   315 &   294 & EDD \\
 KK152    &  82.94+4.19  &  835 &  9.77 &  21 &  8.30 & 135 &   310 &   116 & this work \\
 NGC4517  & 114.82{-}5.32&  972 &  8.36 &  12 &  8.67 & 156 &    76 & {-}12 & EDD \\
 NGC4656  &  84.68+6.02  &  643 &  7.96 &  20 &  9.57 & 143 &   427 &   351 & this work \\
 NGC4631  &  84.21+5.74  &  581 &  7.35 &  20 & 10.09 & 145 &   476 &   417 & EDD \\
 NGC4455  &  92.75+0.19  &  588 &  6.61 &  10 & 10.22 & 163 &   415 &   398 & this work \\
 IC3840   &  95.26+5.02  &  510 &  5.97 &  11 & 10.84 & 164 &   490 &   477 & this work \\
 KK144    &  87.12+1.16  &  452 &  5.99 &  16 & 11.02 & 155 &   562 &   543 & this work \\
 AGC749241&  90.11+3.65  &  418 &  5.62 &  14 & 11.28 & 159 &   587 &   575 & \citet{quinn2014}  \\
 Arp211   &  77.94+6.41  &  484 &  6.14 &  26 & 11.48 & 140 &   586 &   520 & EDD \\
 KDG215   &  97.94+5.26  &  362 &  4.83 &   9 & 11.90 & 167 &   629 &   625 & EDD \\
 KK177    &  95.64+7.54  &  228 &  4.82 &  12 & 11.98 & 163 &   763 &   768 & this work \\
 UA319    & 133.53{-}2.93&  549 &  5.75 &  31 & 12.06 & 135 &   564 &   419 & this work \\
 DDO133   &  84.65+3.58  &  320 &  4.88 &  19 & 12.15 & 153 &   690 &   682 & EDD \\
 DDO126   &  78.93+4.02  &  230 &  4.97 &  25 & 12.31 & 145 &   780 &   805 & EDD \\
 NGC4395  &  82.30+2.73  &  313 &  4.76 &  21 & 12.33 & 151 &   701 &   693 & EDD \\
 NGC4826  &  95.60+6.12  &  363 &  4.41 &  11 & 12.35 & 165 &   631 &   624 & EDD \\
 PGC038685&  78.54+0.54  &  340 &  4.85 &  25 & 12.40 & 146 &   689 &   669 & EDD \\
 UGC07605 &  80.39+3.91  &  316 &  4.74 &  23 & 12.44 & 148 &   705 &   693 & EDD \\
\hline
\end{tabular}
\end{table*}

\addtocounter{table}{-1}
\begin{table*}
\centering
\caption{Continued.}
\label{table:virgogal}
\begin{tabular}{lrrrrrrrrl}
\hline
  Name   &   SGL   SGB &  $V_{LG}$ &  $D$  & $\Theta$ & $D_{VC}$ & $\lambda$ & $V_{VC}^{mi}$ & $V_{VC}^{ma}$ & Ref  \\
         &      deg    &  \kms{}   & Mpc   &  deg  &     Mpc  & deg    &  \kms{}       & \kms{}     \\
\hline
  (1)    &      (2)    &  (3)      & (4)   &  (5)  &     (6)  &   (7)  &    (8)        & (9)   & (10)     \\
\hline
 DDO127   &  78.92+4.31  &  290 &  4.72 &  25 & 12.53 & 146 &   731 &   727 & EDD \\
 LVJ1243+41& 75.60+8.30  &  444 &  4.82 &  29 & 12.66 & 140 &   626 &   541 & this work \\
 IC3687    & 78.42+7.27  &  377 &  4.57 &  26 & 12.71 & 145 &   664 &   620 & EDD \\
 DDO154    & 90.12+6.89  &  354 &  4.04 &  16 & 12.81 & 159 &   649 &   634 & EDD \\
 NGC4244   & 77.72+2.40  &  257 &  4.31 &  26 & 12.90 & 156 &   760 &   760 & EDD \\
\hline
\end{tabular}
\end{table*}

\begin{figure*}
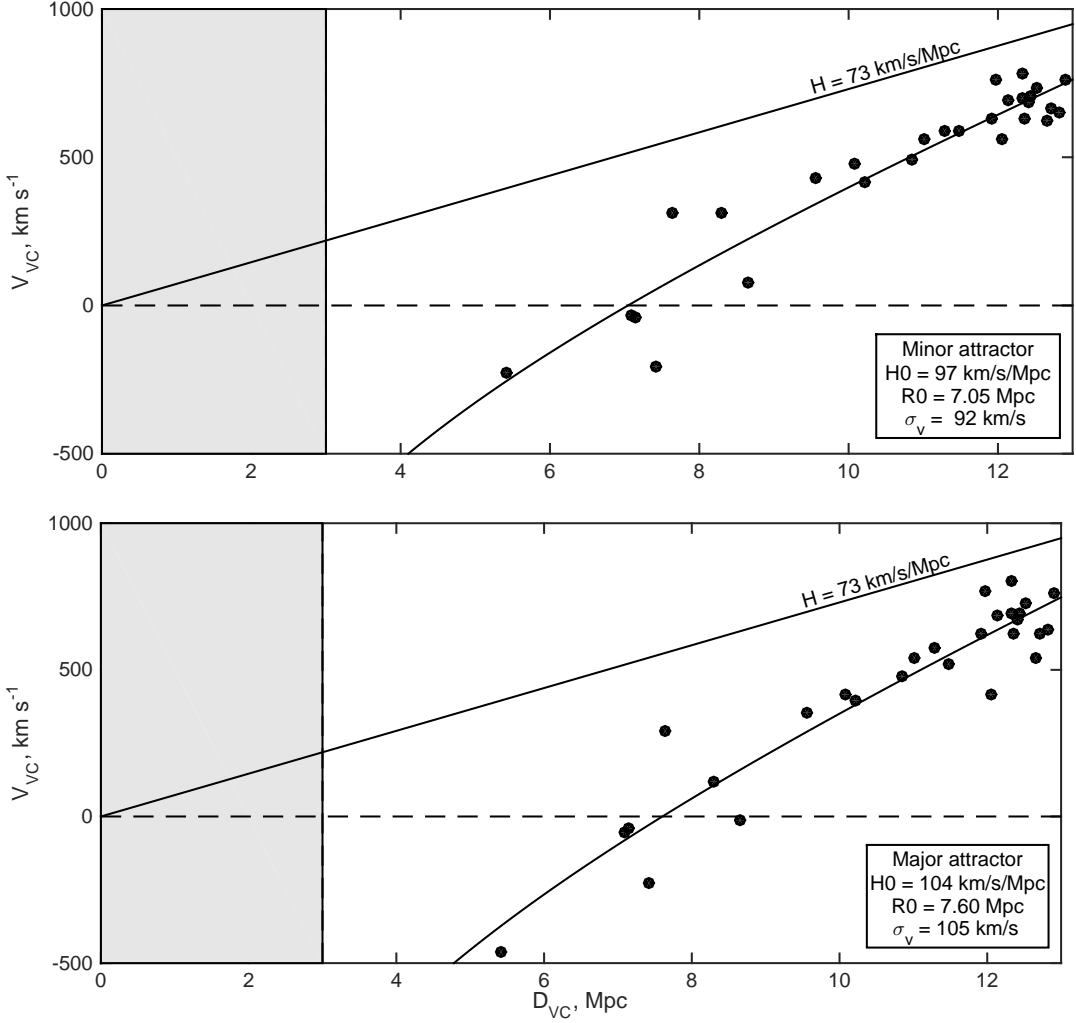

\includegraphics[width=15cm]{Figure4a.eps}
\includegraphics[width=15cm]{Figure4b.eps}
\caption{Hubble diagram "velocity - distance" in the Virgo Cluster frame. Upper 
and lower panels correspond to the cases of "minor" and "major" attractor, respectively.
The area of virial motions is shaded.
}
\label{fig:virgo_distrib}
\end{figure*}

The distribution of 28 galaxies by distances and velocities relative to
the Virgo Cluster center for the cases of minor and major attractors are presented
in the upper and lower panels of the Fig.~\ref{fig:virgo_distrib}. According to \citet{pen2017}, 
the radial velocity profile around a spherically symmetrical overdensity can
be expressed as
$$ V_{VC}(D_{VC}) = H_0\times D_{VC} - H_0\times R_0\times (R_0/D_{VC})^{1/2}, $$
where $R_0$ is the radius of the zero-velocity surface to be found.
The solid line in Fig.~\ref{fig:virgo_distrib} corresponds to this equation with parameters
defined from the least squares method: $R_0 = 7.05$ Mpc, $H_0 = 97$ km\,s$^{-1}$\,Mpc$^{-1}$, 
and dispersion $\sigma_V = 92$ \kms{} for the model of minor attractor, and
$R_0 = 7.60$ Mpc, $H_0 = 104$ km\,s$^{-1}$\,Mpc$^{-1}$, $\sigma_V =105$ \kms{}
for the case of major one. 

To estimate the input of distance errors into the galaxy scatter on the
Fig.~\ref{fig:virgo_distrib} diagrams we adopt their typical distance error of ~5\%. As seen from 
the diagrams, the $R_0$ value is the most sensitive to positions of seven 
galaxies having $D_{VC} < 9$ Mpc. At their average distance from the Milky 
Way of 9.7 Mpc, the average scatter of Virgo-centric distances is 0.57 Mpc.
This corresponds to the Virgo-centric velocity scatter of 78 \kms{} and 90 \kms{}
for the cases of minor and major attractor, respectively. Consequently, we can 
say that the velocity dispersion on the panels of Fig.~\ref{fig:virgo_distrib} is low and contributed 
mostly by galaxy distance errors. The cold Hubble flow around the Virgo Cluster 
may impose constraints on some models of cluster formation.

In the standard cosmological $\Lambda$CDM model, where $\Omega_m$ is the mean
cosmic density of matter, $G$ is the gravitational constant and $H_0$ the Hubble
parameter, the relation between $R_0$ and the total mass of overdensity is
$$ M_T=(\pi^2/8G)\times R^3_0\times H^2_0/f^2(\Omega_m).$$
Here, the dimensionless parameter $f(\Omega_m)$ changes in the range from
1 to 2/3 while varying $\Omega_m$ from 0 to 1. Taking the WMAP parameters:
$\Omega_m =0.24, \Omega_{\lambda} = 0.76$ and $H_0$ = 73 km\,s$^{-1}$\,Mpc$^{-1}$,
we obtain the relation
$$ (M_T/M_{\sun})_{0.24} = 2.12\times10^{12} (R_0/Mpc)^3 $$.
Accordingly, the radius $R_0 = 7.32\pm0.28$ Mpc yields the total mass estimate
for the Virgo cluster $M_T=(8.31\pm0.94)\times10^{14}M_{\sun}$. Under the Planck model parameters 
($\Omega_m =0.315, \Omega_{\lambda} = 0.685$ and $H_0$ = 67.3 km\,s$^{-1}$\,Mpc$^{-1}$),
the total mass estimate drops to $M_T=(7.64\pm0.91)\times10^{14}M_{\sun}$.
This quantity derived from external galaxy motions agrees well
with the internal virial mass estimates: $6.2\times10^{14}M_{\sun}$ \citep{dv60},
$7.5\times10^{14}M_{\sun}$ \citep{tsh84}, $7.2\times10^{14}M_{\sun}$ \citep{gir99}. 
The location of the turnaround radius and mass within it agree in detail with the values 
found by numerical action orbit reconstructions \citep{shaya2017} of $7.3\pm0.3$~Mpc
and $8.3\pm0.3 \times 10^{14}$~\Msun.

\section{Conclusions}

The particular targets of these most recent observations with HST were chosen to clarify 
the situation regarding the transition to infall toward the cluster.  The location of this transition 
provides an accurate estimate of the mass associated with the cluster out to the transition radius.

Of the 20 program targets, 16 distances less than 10 Mpc were successfully measured
with characteristic accuracies of 5\% using the TRGB methodology.  The remaining four galaxies 
are inferred to lie beyond 13 Mpc, in spite of known velocities less than 500~\kms\ for three of them; 
one probably associated with the Virgo core and the other two adding to a curious anomaly
associated with the Coma~I Group.

The new measured distances add to pre-existing distances in an analysis of the velocity field
in the proximity of the transition between Hubble expansion and Virgo infall.
Within a reasonable range of assumptions, the Virgo turnaround radius is located at
$R_0 = 7.3\pm0.3$~Mpc and the mass inside this radius is $8.3\pm0.9 \times 10^{14}$~\Msun.
These values are in close agreement with measures from numerical action orbit reconstructions 
based on a much larger compendium of distances although lacking the 16 important new distances 
reported here.

\acknowledgments
This research is based on observations made with the NASA/ESA
Hubble Space Telescope, program GO-14636, with data archive at
the Space Telescope Science Institute. STScI is operated by the
Association of Universities for Research in Astronomy, Inc. under
NASA contract NAS 5-26555.
The authors thank the anonymous referee for a thorough
examination of the manuscript and for useful comments and
suggestions to improve the text. 
Support for the program
GO-14636 was provided by NASA through a grant from the
Space Telescope Science Institute, which is operated by the
Association of Universities for Research in Astronomy, Inc.,
under NASA contract NAS 5-26555.
I.K. and L.M. acknowledge the support by the Russian Science Foundation grant 14--12--0096
and by RFBR grant 18-02-00005. The analysis of the galaxy distribution is supported by 
the Russian Science Foundation grant 14--12--0096.
This research has made use of the NASA/IPAC Extragalactic Database (NED),
which is operated by the Jet Propulsion Laboratory, California Institute of Technology,
under contract with the National Aeronautics and Space Administration.
We acknowledge the usage of the HyperLeda database (http://leda.univ-lyon1.fr).
Help in the fitting Fig.~\ref{fig:SGBL} and Fig.~\ref{fig:virgo_distrib} 
by Olga Kashibadze is greatly appreciated.

\facility{HST (ACS)}

\software{
DOLPHOT \citep{d2000},
trgbtool \citep{makarov06}
}

\bibliography{virgo}   


\end{document}